# Effect of Collective Molecular Reorientations on Brownian Motion of Colloids in Nematic Liquid Crystal

T. Turiv[1,2], I. Lazo[2], A. Brodin[1,3], B. I. Lev[4], V. Reiffenrath[5], V. G. Nazarenko[1], and O. D. Lavrentovich[2]*

[1]Institute of Physics NASU, prospect Nauky 46, Kyiv, 03039, Ukraine.
[2]Liquid Crystal Institute, Kent State University, Kent, OH 44242.
[3]National Technical University of Ukraine "KPI", prosp. Peremogy 37, Kyiv 03056, Ukraine.
[4]Bogolyubov Institute for Theoretical Physics NASU, 14-b, Metrolohichna Str., Kyiv 03680, Ukraine.
[5]MERCK KGaA, Liquid Crystals Division, 64271 Darmstadt, Germany.

*To whom correspondence should be addressed: olavrent@kent.edu

**In the simplest realization of Brownian motion, a colloidal sphere moves randomly in an isotropic fluid; its mean squared displacement (MSD) grows linearly with time $\tau$. Brownian motion in an orientationally ordered fluid, a nematic, is anisotropic, with the MSD being larger along the axis of molecular orientation, called the director. We show that at short time scales, the anisotropic diffusion in a nematic becomes also anomalous, with the MSD growing slower (subdiffusion) and faster (superdiffusion) than $\tau$. The anomalous diffusion occurs at time scales that correspond to the relaxation times of director deformations around the sphere. Once the nematic melts, the diffusion becomes normal and isotropic. The experiment shows that the deformations and fluctuations of long-range orientational order profoundly influence diffusive regimes.**

Random displacements of a small particle in a fluid are controlled by kinetic energy dissipation (*1*). The mean displacement is zero, but the average mean squared displacement (MSD) is finite, growing linearly with the time lag $\tau$ (*2*), $\langle \Delta \mathbf{r}^2(\tau) \rangle = 6D\tau$, where $D$ is the translational diffusion coefficient. Brownian particles in complex fluids may exhibit an anomalous behavior, $\langle \Delta \mathbf{r}^2(\tau) \rangle \propto \tau^\alpha$, with the exponent $\alpha$ either smaller than 1 (subdiffusion) or larger than 1 (superdiffusion). Subdiffusion is observed in polymer (*3*) and F-actin networks (*4*), in surfactant dispersions (*5*); superdiffusion occurs in concentrated suspensions of swimming bacteria (*6*) and dispersions of polymer-like micelles (*7-10*). The diffusion regimes should reflect the properties of the host medium (*11*), one of which is often a local or long-range orientational order of molecules.

The simplest orientationally ordered fluid is a uniaxial nematic, in which the average orientation of molecules is described by a unit director $\hat{\mathbf{n}}$. Because of different effective viscosities $\eta_\parallel \neq \eta_\perp$ for motion parallel and perpendicular to $\hat{\mathbf{n}}$, Brownian motion becomes anisotropic, with two coefficients $D_\parallel$ and $D_\perp$ (*12-19*). The anisotropic diffusion characterized experimentally so far for nematics at relatively long time lags $\tau$ remains "normal", with $\alpha=1$ (*14-19*). In some cases, anomalous diffusion has been also observed, but it was attributed to features other than the orientational order, such as bacterial activity (*6*), size distribution of building units (*7*), spatial modulation of hydrophobic and hydrophilic regions (*8*), bending rigidity of the molecular aggregates (*8*), fluctuations of concentration (*9*), director distortions around the dye molecules (*20*), etc. In this work, we demonstrate that the nematic orientational order and its fluctuations alone can cause the diffusion to be both anisotropic and anomalous.

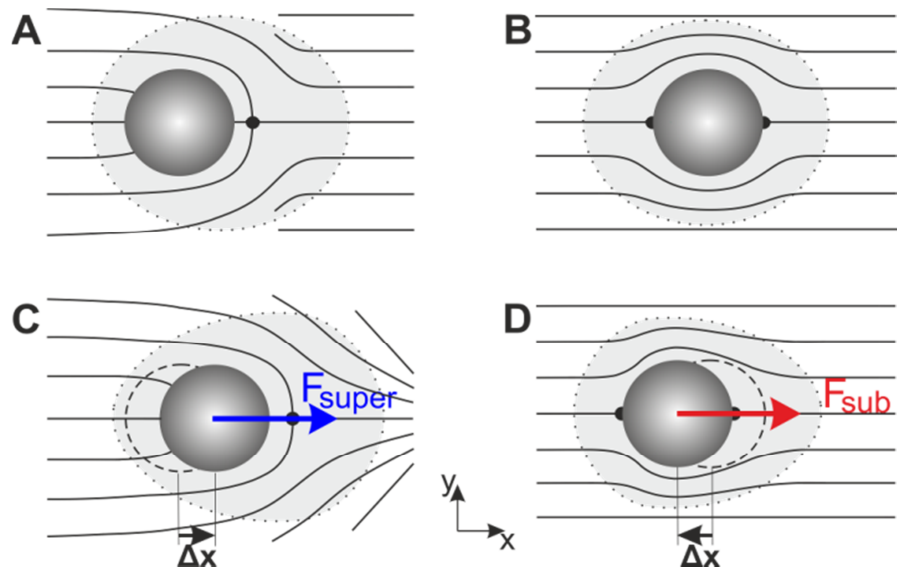

**Fig. 1. Director deformations around a sphere and mechanisms of super- and subdiffusion in nematic.** (**A**) Equilibrium director for normal and (**B**) tangential surface anchoring. (**C**) Elastic force $\mathbf{F}_{super}$ caused by a fluctuative splay on the right hand side, moves the sphere towards the splay. (**D**) Restoring force $\mathbf{F}_{sub}$ responds to displacement $\Delta x$ that creates stronger director gradients on the left hand side.

We used silica spheres of diameter $d = (1.6-10)$ μm with two types of surface alignment (*21*) that set either perpendicular, Fig. 1A or tangential, Fig. 1B, orientation of $\hat{\mathbf{n}}$. The spheres are placed in a cell with uniform director alignment $\hat{\mathbf{n}}_0$ set by unidirectionally treated bounding plates (*21*). The director distortions around the particle (*22*) cause repulsion from the bounding substrates, so that the colloids levitate in the nematic bulk (*23*), Fig. S1. To minimize experimental errors caused by birefringence, we use the nematic IS-8200 with ultra-low birefringence $\Delta n = 0.0015$, Fig. S2 (*24*).

The typical measured MSD vs $\tau$ dependencies are presented in Fig. 2 for perpendicular anchoring and in Fig. S5 for tangential anchoring. In the isotropic phase, $T = 60°C$, diffusion is normal with $D = 9.2 \times 10^{-16}$ m$^2$/s. In the nematic, at $T = 50°C$, diffusion becomes anisotropic. At relatively long time scales, $\tau > (20-40)$ s, this anisotropic diffusion is normal, as both MSD components, measured for displacements parallel and perpendicular to $\hat{\mathbf{n}}_0$, grow linearly with $\tau$. We find $D_\parallel = 1.9 \times 10^{-16}$ m$^2$/s and $D_\perp = 1.4 \times 10^{-16}$ m$^2$/s for the normally anchored spheres and $D_\parallel = 2.2 \times 10^{-16}$ m$^2$/s, $D_\perp = 1.3 \times 10^{-16}$ m$^2$/s for tangential anchoring. In all these cases $\alpha = 1.0 \pm 0.03$. At the times shorter than (20-40) s, the dependence MSD vs $\tau$ becomes nonlinear, Fig. 2A and S5A, signaling anomalous regimes.

To obtain a better insight into the different diffusion regimes, we calculated the velocity autocorrelation function (VACF) $C_{v\parallel}(\tau) = \langle v_x(\tau)v_x(0)\rangle$ (*25-26*), where $v_x$ is the velocity of the particle along the $x$-axis, and a similar quantity $C_{v\perp}(\tau) = \langle v_y(\tau)v_y(0)\rangle$ for $y$-direction. For the isotropic melt, Fig. 2D, S5D, and for the long time scales in the nematic, $C_{v\parallel}(\tau)$ and $C_{v\perp}(\tau)$ are close to zero, Fig. 2B-2C, S5B-S5C, as it should be when $\alpha = 1$. However, when the time steps for the nematic are within a certain interval $\tau_{\text{sup}} < \tau < \tau_{\text{sub}}$, both $C_{v\parallel}(\tau)$ and $C_{v\perp}(\tau)$ become negative, indicating subdiffusion. At yet shorter times $\tau < \tau_{\text{sub}}$, $C_{v\parallel}(\tau)$ and $C_{v\perp}(\tau)$ are positive, reflecting superdiffusion.

Experimentally, one determines $\tau_{sup}$ as a time point at which the VACF changes its sign, and $\tau_{sub}^{min}$, at which the function reaches its minimum; $\tau_{sub}$ is determined approximately when the deviation from zero exceeds $10^{-17}$ m$^2$/s$^2$, a typical scatter of VACF data in Fig. 2B, 2C. We find that $\tau_{sup}$ and $\tau_{sub}^{min}$ increase quadratically with the sphere's diameter, Fig. 3.

The dynamic complexity of the dynamic director environment around the Brownian particles and finite span (in terms of time lags) make each of the anomalous regimes unlikely to follow a single power law with a fixed $\alpha$. Qualitatively, the existence of slopes in the MSD vs $\tau$ dependencies that are smaller than 1 (for $\tau_{sup} < \tau < \tau_{sub}$) and larger than 1 (for $\tau < \tau_{sup}$) is seen in the log-log plot shown in Fig.S13D, S14 and S15 (*21*). If one does oversiplifies the situation and assumes a single power law, MSD $\propto \tau^\alpha$ for the $\tau$ limits identified above, then the best fit for the subdiffusive domain $\tau_{sup} < \tau < \tau_{sub}$ in IS-8200 is achieved for $\alpha_\parallel = 0.35 \pm 0.01$ and $\alpha_\perp = 0.30 \pm 0.01$; for $\tau < \tau_{sup}$, one obtains $\alpha_\parallel = 1.32 \pm 0.01$ and $\alpha_\perp = 1.20 \pm 0.01$.

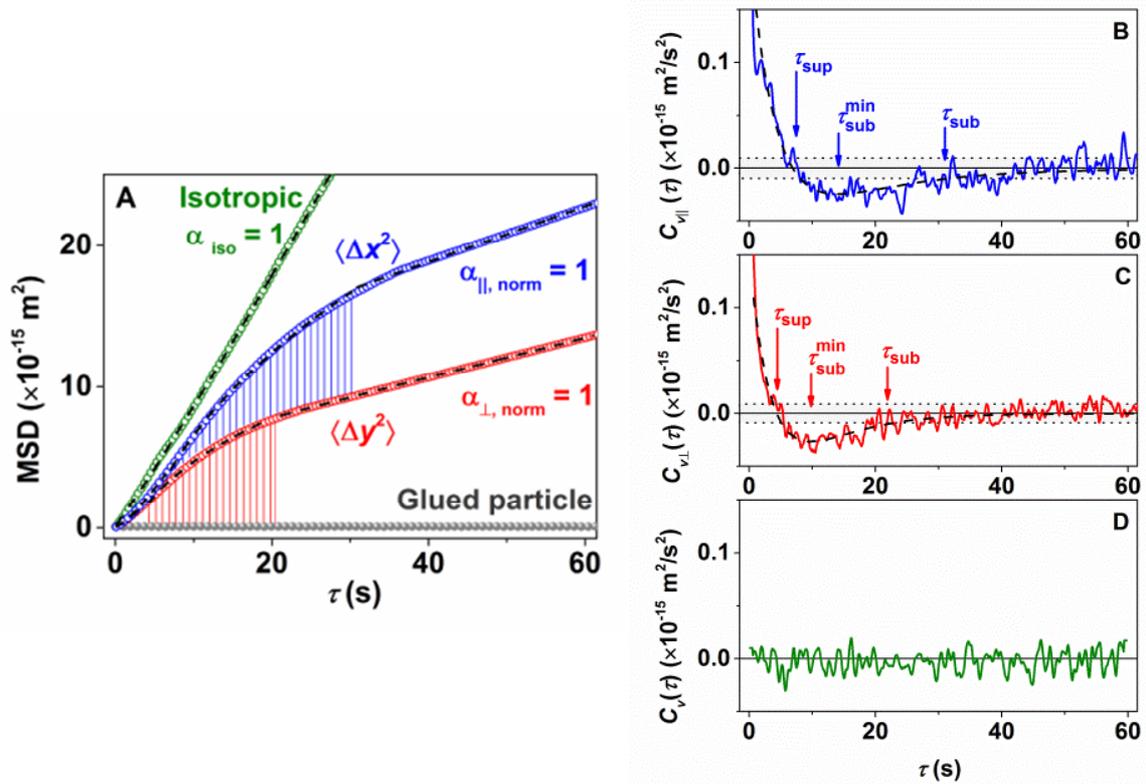

**Fig. 2. MSDs and velocity autocorrelation functions of 5 μm silica spheres in IS-8200.** (**A**) MSD versus time lag for normally anchored sphere in the isotropic ($T = 60°C$) and nematic ($T = 50°C$) phases of IS-8200, in the direction parallel ($x$) and perpendicular ($y$) to the overall director $\hat{\mathbf{n}}_0$. Vertically dashed pattern labels the subdiffusion domains. The bottom curve represents apparent MSD of a particle glued to the cell substrate (*27*). Cell thickness 50 μm. (**B**) VACF for the normally anchored sphere moving parallel to $\hat{\mathbf{n}}_0$ and (**C**) perpendicular to $\hat{\mathbf{n}}_0$. (**D**) The same for the isotropic phase. Dashed lines in (**A**-**C**) are least-square fits. Levels of noise in the VACF data for the linear domain are shown in (**B**) and (**C**) as dotted horizontal lines.

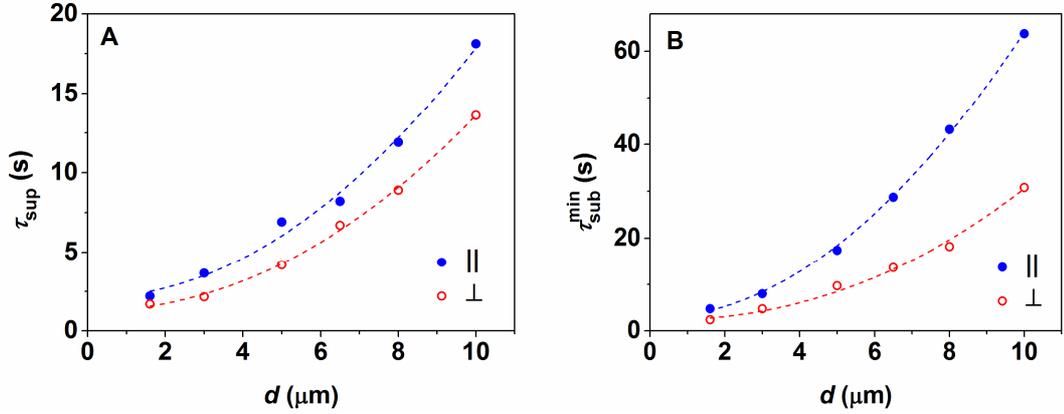

**Fig. 3. Characteristic times of anomalous diffusion vs sphere diameter** $d$. (**A**) $\tau_{sup}$ and (**B**) $\tau_{sub}^{min}$, measured for normally anchored spheres in the nematic IS-8200 paralell and perpendicular to $\hat{\mathbf{n}}_0$. Cell thickness 50 µm. $T = 50°C$. Dashed lines are quadratic fits.

Anomalous diffusion is also featured in the probability distribution of particle displacements. We calculate the probability distribution function $G(\Delta x, \tau)$ for displacements parallel to $\hat{\mathbf{n}}_0$, Fig. 4, and $G(\Delta y, \tau)$ for the perpendicular component (Fig. S16). Both functions show a distinct behavior within different time scales. For the longest time lags, $\tau = 40$ s, $G(\Delta x, \tau)$ coincides with the Gaussian fitting $G_G(\Delta x)$. Such a behavior is characteristic of normal diffusion (*14*, *18*). For the intermediate time lags, $\tau = 10$ s, only the central part of $G(\Delta x, \tau)$ (small displacements) is approximately Gaussian. The probability of finding a large displacement at these time lags is noticeably lower as compared to the Gaussian distribution. This behavior correlates with the idea of subdiffusion, as the particle remains close to its original position. Finally, for the shortest time lags, $\tau = 1$ s, Fig. 4, large displacements are more probable as compared to the normal diffusion; this behavior is indicative of superdiffusion.

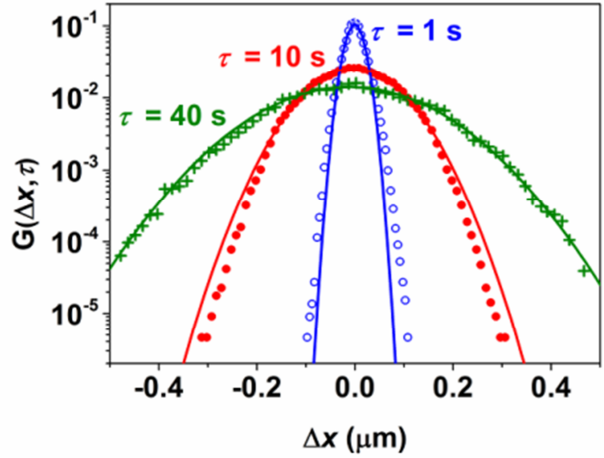

**Fig. 4. Probability distribution of particle displacements parallel to the overall director, for different time lags, 1 s, 10 s, and 40 s.** 5 µm silica spheres with normal anchoring in IS-8200, $T = 50°C$. The solid lines are Gaussian fits. The data are obtained from a sample of 200,000 single particle's trajectory steps.

The nematic IS-8200 is not unique in terms of the observed effects. In an appropriate range of timescales, subdiffusion is observed also in the thermotropic 5CB (Fig. S11, S14) and in the lyotropic chromonic LC DSCG (Fig. S12, S15).

The experiments demonstrate that the effect of orientationally ordered environment on Brownian motion is very profound, causing, in addition to anisotropy, anomalous super- and subdiffusion. These two regimes are observed at relatively short time scales that vary with the material, type of anchoring at the particle's surface, size, and displacement direction with respect to $\hat{\mathbf{n}}_0$. Above these short time scales, the diffusion becomes normal (but still anisotropic).

The current models of Brownian motion in a nematic (*12-14,18*) consider the director field around the particle as stationary. The predicted diffusion is normal albeit anisotropic (*12-18*). We attribute the observed anomalous diffusion to the coupling of the sphere's displacements to the director field and its fluctuations, i.e. to an intrinsic memory (*28*) of the system associated with the orientational order, as discussed below.

The director field $\hat{\mathbf{n}}(\mathbf{r},t)$ is coupled to the velocity field $\mathbf{v}(\mathbf{r},t)$ of the nematic. Both $\hat{\mathbf{n}}(\mathbf{r},t)$ and $\mathbf{v}(\mathbf{r},t)$ are perturbed by the particle and by the director fluctuations. A perturbation of the director around a colloid of diameter $d$ relaxes within a characteristic time (29) $\tau_d \sim \beta^2 \eta_{eff} d^2 / K$, where $\beta$ is a numerical coefficient of the order of 1 that describes the lengthscale $\sim \beta d$ of deformations (30), $\eta_{eff}$ and $K$ are the effective viscosity and elastic constant. For IS-8200, $\eta_{eff}/K \approx 10^{11}$ s/m$^2$ (Fig. S7-S10) (21) and thus $\tau_d$ is in the range (depending on $d$) $\tau_d = (0.1-10)$ s, being thus much larger than the time $\sim \rho d^2/\eta \sim 1\,\mu$s needed by the perturbed fluid of density $\rho$ and viscosity $\eta \sim 0.1$ Pa·s to flow over the distance $d$ (29). It means that the isotropic hydrodynamic memory is not the reason of anomalous diffusion, consistent with data for the isotropic melt, Fig. 2A, 2D, S5A, and S5D.

The coupling of director and displacements that leads to subdiffusion at the time scales $\tau_{sub}, \tau_{sub}^{min} \sim \tau_d$ can be illustrated with a tangentially anchored sphere, Fig. 1B. A fluctuative displacement of the sphere, say, to its left by a distance $\Delta \mathbf{x}$, Fig. 1D, temporarily increases the elastic energy density on the left hand side and decreases it on the right hand side. The difference creates a restoring elastic force $\mathbf{F}_{sub} \sim -K\Delta\mathbf{x}/d$ that slows down the diffusive motion.

The case of superdiffusion is sketched in Fig. 1C for a normally anchored sphere that finds itself near a fluctuative splay of the director. The particle experiences an attractive or repulsive force towards the splay n, $\mathbf{F}_{super} = -\nabla U$, where $U = -4\pi K \int \mathbf{P}\cdot\mathbf{n}(\nabla\cdot\mathbf{n})dV$ and $\mathbf{P}$ is the elastic dipole of the sphere, $P \propto d^2$ (30). This force leads to superdiffusive motion during the lifetime of splay fluctuation.

The memory span in the examples above is determined mostly by $\tau_d$. For $\tau \ll \tau_d$, the influence of perturbations with wavelength much shorter than $d$ averages to zero. The lifespan of fluctuations is also limited from above, by the cell thickness $h$, $\tau_h \sim \eta_{eff} h^2/\pi^2 K$. For $h \gg d$, the thickness dependence is expected to be weak and dependent on how far from the boundary the sphere levitates (23); the latter is determined not only by $h$ but also by densities, elastic and surface anchoring parameters.

For $d = 5$ µm in an IS-8200 cell of thickness 50 µm, we estimate $\tau_d \approx 2.5$ s (assuming $\beta = 1$ and $\eta_{eff}/K = 10^{11}$ s/m$^2$) and $\tau_h \approx 30$ s. These values correlate well with the experimentally found $\tau_{sup} = (4-10)$ s and $\tau_{sub} = (20-42)$ s, Fig. 2 and Table S2. The quadratic dependence of $\tau_{sup}$ and $\tau_{sub}^{min}$ on $d$, Fig. 3, is another strong indication of the director relaxation mechanism.

The considerations above are not restricted to the specific examples in Fig. 1C, 1D. For example, a tangentially anchored particle might be attracted by fluctuative deformation in its neighbourhood, while displacements of a normally anchored sphere would lead to asymmetric director distortions and a subdiffusive slowing-down. Universality of director distortions in terms of attraction/repulsion of colloids are already known for stationary director fields (*31*).

To conclude, our work demonstrates that the orientational order in a nematic liquid crystal causes a profound effect on Brownian motion of a small spherical particle and results in anisotropic subdiffusion and superdiffusion. These two are different from the normal diffusion observed in simple isotropic fluids and from anisotropic but linear behavior of MSD established for long time scales in liquid crystals. We observe that as the time scale $\tau$ decreases, the normal diffusion is replaced by subdiffusion and then by superdiffusion. All three regimes are anisotropic, with MSDs in the direction parallel to the overall director and perpendicular to it being different. The characteristic times at which the anomalous regimes are observed, are within the range of relaxation times of the director distortions around the particle and director fluctuations. Although this study dealt with standard liquid crystals, the observed anomalous diffusion is expected to arise in any dispersive environment with orientational order.

# Supplementary Materials

## Materials and methods

### Materials

We used silica spheres of diameter $d = 1.6, 3, 5, 6.5, 8$, and $10\ \mu\text{m}$ (Bangs Laboratories), with two types of surface alignment: perpendicular, Fig. 1A, and tangential to the particle's surface, Fig. 1B. In the first case, the spheres were functionalized with octadecyl-dimethyl(3-trimethoxysilylpropyl) ammonium chloride (DMOAP) (*32*). In the second case, the particles were left untreated. The overall uniform orientation of the nematic was set by two glass plates covered with rubbed polyimide PI-2555 (Nissan Chemicals) that produces a unidirectional planar alignment. The locally distorted director around the spheres must smoothly transform into the uniform field far away from it. The resulting equilibrium director is either of a dipolar type (with a point defect-hyperbolic hedgehog accompanying the sphere) for normal anchoring, Fig. 1A, or of a quadrupolar type with two surface defects-boojums at the poles, in the tangential anchoring case, Fig. 1B (*30*). The director distortions around the particle (*22*) cause repulsion from the bounding substrates, so that the colloid levitates in the bulk at some height determined by the balance of gravity and elastic forces (*23*), Fig. S1.

To minimize the image drifting in highly birefringent medium we use the nematic IS-8200 (*24*), Fig. S2 and Table S1, with ultra-low birefringence $\Delta n = 0.0015$ (at $\lambda = 520$ nm) that is two orders of magnitude smaller than birefringence of a standard nematic pentylcyanobiphenyl (5CB) with $\Delta n = 0.17$. For $R \sim 5\ \mu\text{m}$, the image drift in IS-8200 is less than $\sim 10$ nm. We also studied 5CB and a lyotropic nematic, formed by water solutions of disodium chromoglycate (DSCG) with a relatively low negative birefringence $\Delta n = -0.015$ (at $\lambda = 520$ nm).

### Methods

#### Particle Tracking

We used high-speed video camera MotionBlitz EoSens mini1 (Microtron GmbH) mounted on an inverted microscope Nikon ECLIPSE TE2000-U with a $100\times$ 1.3 N.A. immersion objective to trace the particle trajectories. To control the temperature, we used the heating stage Linkam LTS120 (accuracy $0.1°\text{C}$). Digital images of isolated colloidal particles, captured at a maximum frame rate of 2400 fps (time resolution 0.4 ms), were analyzed to find the coordinates (*x*, *y*) of the particle's center using the intensity-weighted algorithms (*33*).

Sequences of images were tracked continuously for more than 100 s. As discussed in the main text, tracking of the particle's position in a LC is complicated by birefringence of the material. To establish the limit of accuracy in measurements of particle's positions that depends on birefringence, we used particles immobilized (by a Norland adhesive) at the bottom plate of the cell filled with the LC. First, we determined the coordinates of immobilized particles in an empty cell, as $x = \dfrac{\sum_i I_i x_i}{\sum_i I_i}$ and $y = \dfrac{\sum_i I_i y_i}{\sum_i I_i}$, where $x_i$ and $y_i$ are the coordinates of the $i$-th pixel, $I_i$ is its intensity, Fig. S3. Within each frame with 8-bit gray scale (0-255 arbitrary units of intensity), only the pixels with intensity higher than a certain threshold value $I_{i,threshold}$ were taken into account; $I_{i,threshold}$ was determined from the condition that the mean square displacement of the immobilized particle is minimum, Figure S3E.

The experiments with the empty cells were followed by experiments with cells filled with five different fluids: three types of a nematic LC and two isotropic fluids (water and glycerol). In all cases, we determined the MSD of particles immobilized at the bottom plate, to maximize the effect of the medium on the measured MSD. The probing light beam travelled through the entire thickness of the cell. The apparent mean square displacement of the immobilized particles versus time lag is shown in Fig. S4. The apparent displacements represent a cumulative effect of errors in measuring the particle's position caused by the optical system of the microscope, vibrations and birefringence. It grows with birefringence of the material, being the largest for the nematic pentylcyanobiphenyl (5CB) with the highest birefringence (~0.2). In all cases, the apparent MSD in the time range of interest was about $10^{-16}$ m$^2$ or less; these values are about 100 times smaller than the MSD of free spheres experiencing anomalous diffusion described in the main text.

Because of the long range elastic interaction between particles in a nematic, we used only very low concentrations of the colloidal particles, so that the distance between them is much larger (by a factor of 50 at least) than their diameter. The trajectories were measured for individual particles in separate samples. As an example, we show data for cells of the same thickness 50 µm containing similar 5 µm spheres with homeotropic alignment, Fig. S6. The overall behavior of trajectories for different spheres is the same, with some variations of the characteristic times; we attribute these differences to variations of anchoring conditions on the particles surface.

**Probability distribution**

The difference in diffusion regimes is also examined by measuring the probability distribution of particle displacements. We calculate the probability distribution function $G(\Delta x, \tau) = \frac{1}{N} \sum_{i=1}^{N} \delta(\Delta x - |x_i(0) - x_i(\tau)|)$ for displacements parallel to $\hat{\mathbf{n}}_0$, Fig. 4, and a similar function $G(\Delta y, \tau)$ for the perpendicular component (*34*), Fig. S16. Here $N$ is the total number of particle's displacements extracted from the traced trajectory with a time step $\tau$; $\delta$ is the Dirac delta function. We determined $G(\Delta x, \tau)$ and $G(\Delta y, \tau)$ for $d = 5$ μm spheres with normal surface anchoring. The data for a fixed $\tau$ are then compared to the Gaussian probability distribution functions, such as $G_G(\Delta x) = \frac{A}{\sigma\sqrt{2\pi}} \exp\left[-\frac{(\Delta x)^2}{2\sigma^2}\right]$, Fig. 4, where $\sigma$ is the standard deviation of the distribution and $A$ is the normalization constant (and a similar function $G_G(\Delta y)$). The fitting was performed for three different time lags that correspond to the different diffusion regimes suggested by MSD and $C_v(\tau)$ measurements, Fig. 2A-2C.

**Connection of MSD and velocity autocorrelation function**

The MSD along, say, the *x*-axis is expressed through the velocity autocorrelation function as follows (*25*, *26*):

$$\langle \Delta x^2(t) \rangle = 2 \int_0^t dt' \int_0^{t'} C_{V_x}(t'') dt''.$$

Conversely, $C_{V_x}(t) = \frac{1}{2} \frac{d^2 \langle \Delta x^2(t) \rangle}{dt^2}$. Thus, normal diffusion with $\langle \Delta x^2(t) \rangle \propto t$ would lead to $C_V(t) = 0$. If $C_V(t)$ is non-zero, its sign distinguishes subdiffusion ($C_V(t) < 0$, the MSD grows slower than $t$) from superdiffusion ($C_V(t) > 0$, the MSD grows faster than $t$).

**SOM Text**

**1. Director Relaxation Times of IS-8200**

To evaluate the relaxation time of director in IS-8200, we used the electric field-induced Frederiks transition (*29*). Since the dielectric anisotropy of IS-8200 is negative, the simplest approach is to explore the director reorientation in the geometry of bend Frederiks effect (*29*). In the field-free state, the director is aligned perpendicularly to the two glass plates. We used SE-1211 polyimide (Nissan Chemical Industries, Ltd.) for this homeotropic alignment; the polymer layers were rubbed in order to pre-select the direction of director tilt when the electric field is applied to the conductive indium tin oxide (ITO) electrodes deposited at the glass plates beneath the polymer layers. We used cells with gaps of thickness 10, 20, 30, 50, and 75 μm.

The cell is kept in the heating stage HS-1 (Instec Inc.) at 50°C and observed between two crossed polarizers aligned at 45 degrees to the rubbing direction. As the voltage exceeds some threshold $V_{th}$, the director deviates from the vertical orientation and light transmission through the cell increases, Fig. S7. We find $V_{th}$ = 4.7 V for 20 μm cell.

By raising the AC voltage slightly above $V_{th}$ and switching it off, we monitor the relaxation of the optical signal and fit it with the exponential function to determine the relaxation time of bend deformation, Fig. S8. The thickness dependence of relaxation time is shown in Fig. S9; as expected, it follows a quadratic dependence. The ratio $\eta_{eff} / K = 5.8 \times 10^{10}$ s/m$^2$ is estimated from (*29*) as $\frac{\eta_{eff}}{K} = \frac{\tau_{relax} \pi^2}{h^2}$.

We performed a similar experiment for twist Frederiks transition in a planar cell; planar alignment was achieved by oblique evaporation of SiO$_x$. The electric field was applied in the plane of the cell. The dynamics of transmitted light intensity is shown in Fig. S10. The deduced effective value in this case is $\eta_{eff} / K = 16 \times 10^{10}$ s/m$^2$. The typical order-of-magnitude estimate is thus $\eta_{eff} / K \approx 10^{11}$ s/m$^2$.

**2. Diffusion of the particles in 5CB**

In the classic nematic 5CB, the MSD dependencies suggest subdiffusion with $\alpha_{\parallel} = 0.38$ and $\alpha_{\perp} = 0.36$, but only when the time lags are shorter than $\tau_{sub}^{\parallel}$ = 8.4 ms and $\tau_{sub}^{\perp}$ = 6.6 ms, Fig.

S11. The data are presented for 5 µm particles with normal anchoring at $24°C$. At longer times, the diffusion is normal and anisotropic with $D_\parallel = 2.7\times 10^{-15}$ m$^2$/s$^{-1}$ and $D_\perp = 1.3\times 10^{-15}$ m$^2$/s$^{-1}$; the effective viscosities are $\eta_\parallel = 63.5$ mPa s and $\eta_\perp = 90.0$ mPa s. These values fall within the range of the Mieşowicz viscosities known for 5CB (42, 22, and 118 mPa s (*35*)) and are close to the values $\eta_\parallel = 52.5$ mPa s and $\eta_\perp = 86.4$ mPa s, obtained in Ref. (*36*). In the isotropic ($40°C$) phase, the diffusion is normal and isotropic, with $D = 2.9\times 10^{-15}$ m$^2$/s$^{-1}$.

## 3. Diffusion of the particles in lyotropic chromonic liquid crystal DSCG

The experiments on Brownian motion were repeated for the lyotropic chromonic LC disodium chromoglycate (DSCG) (*37*) with $\Delta n = 0.015$ at $\lambda = 520$ nm (*38*). The nematic phase in the water solution with 13 wt % concentration of DSCG exists at room temperature; it melts into an isotropic phase at $29°C$. The elastic constants at the studied temperature and concentration are as follows: splay $K_{11} = 4.3$ pN, twist $K_{22} = 0.7$ pN, and bend $K_{33} = 23$ pN (*39*); the rotational viscosity is estimated as $\gamma_1 \sim 20$ kg m$^{-1}$s$^{-1}$ from the time of director relaxation in a magnetically realigned cell. Subdiffusive behavior is observed again, with $\tau_{sub}^\parallel = 0.37$ s, $\tau_{sub}^\perp = 0.46$ s, $\alpha_\parallel = 0.4$ and $\alpha_\perp = 0.53$, Fig. S12. At longer times, diffusion is normal, with $D_\parallel = 6.8\times 10^{-17}$ m$^2$s$^{-1}$ and $D_\perp = 4.5\times 10^{-17}$ m$^2$s$^{-1}$. In the isotropic phase, Fig. S12A, diffusion is normal and isotropic $D_{iso} = 3.2\times 10^{-14}$ m$^2$s$^{-1}$.

## 4. Scaling exponents and MSD in log-log coordinates

The exponents in the power laws presented in the main text for 5 µm silica particle in IS-8200 were obtained from fitting the data by a single power law in each othe domains, namely, normal domain ($\tau > \tau_{sub}$), Fig. S13A, subdiffusive domain ($\tau_{sup} < \tau < \tau_{sub}$), Fig.S13B, and superdiffusive domain ($\tau < \tau_{sup}$, down to 0.1 s), Fig.S13C. To help in visualization of data in terms of power laws, we present the log-log plots of MSD vs time lag for 5 µm silica particles diffusing in the nematic phases of IS-8200 (Fig.S13D), 5CB (Fig.S14), and chromonic DSCG (Fig.S15). Although the plots in Figs. S13-S15 do show the deviations from the classic normal diffusion regimes when the time lags become shorter, they do not offer a straightforward protocol to extract the values of $\alpha$'s and $\tau$'s (*27*).

**Supporting Figures**

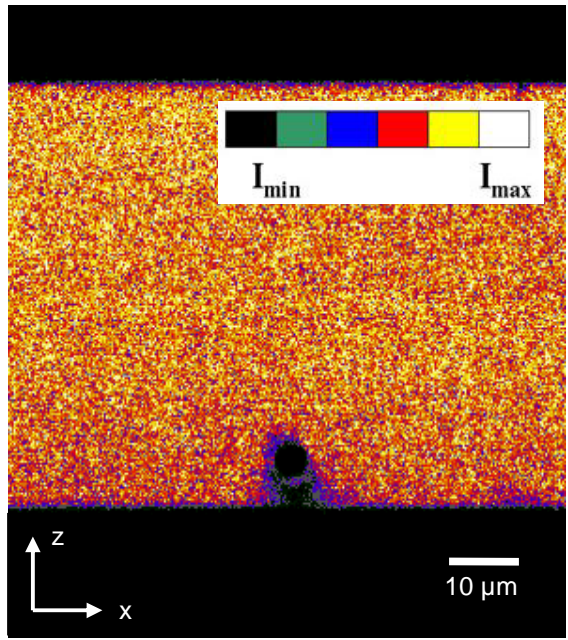

**Fig. S1. Levitation of a sphere in nematic LC IS-8200.** Fluorescent confocal polarizing microscopy texture of a vertical cross-section of a nematic cell of thickness 50 μm filled with IS-8200 and a small amount (< 0.01wt%) of fluorescent dye *n, n'-bis* (2,5-di-tert-butylphenyl)-3,4,9,10-perylen-edicarboximide (BTBP). The glass sphere of diameter 5 μm levitates at the height about 5 μm from the bottom plate. The shadow below the sphere results from light scattering.

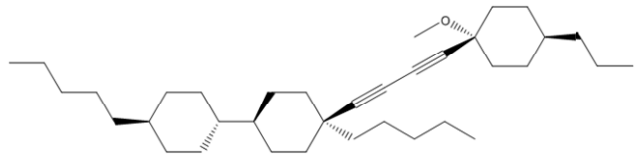

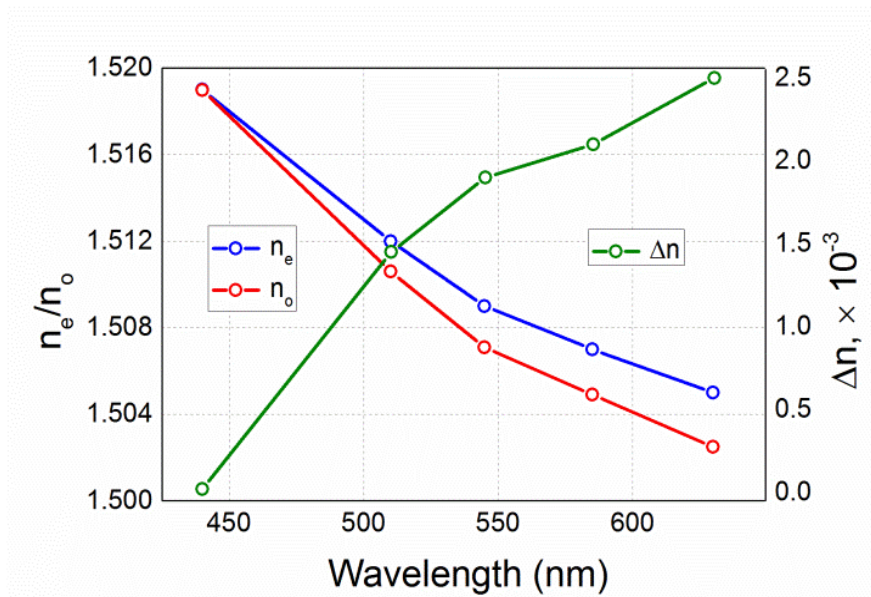

**Fig. S2. Nematic liquid crystal IS-8200.** (**A**) Chemical structure; (**B**) wavelength dependence of the refractive indices and birefringence $\Delta n$.

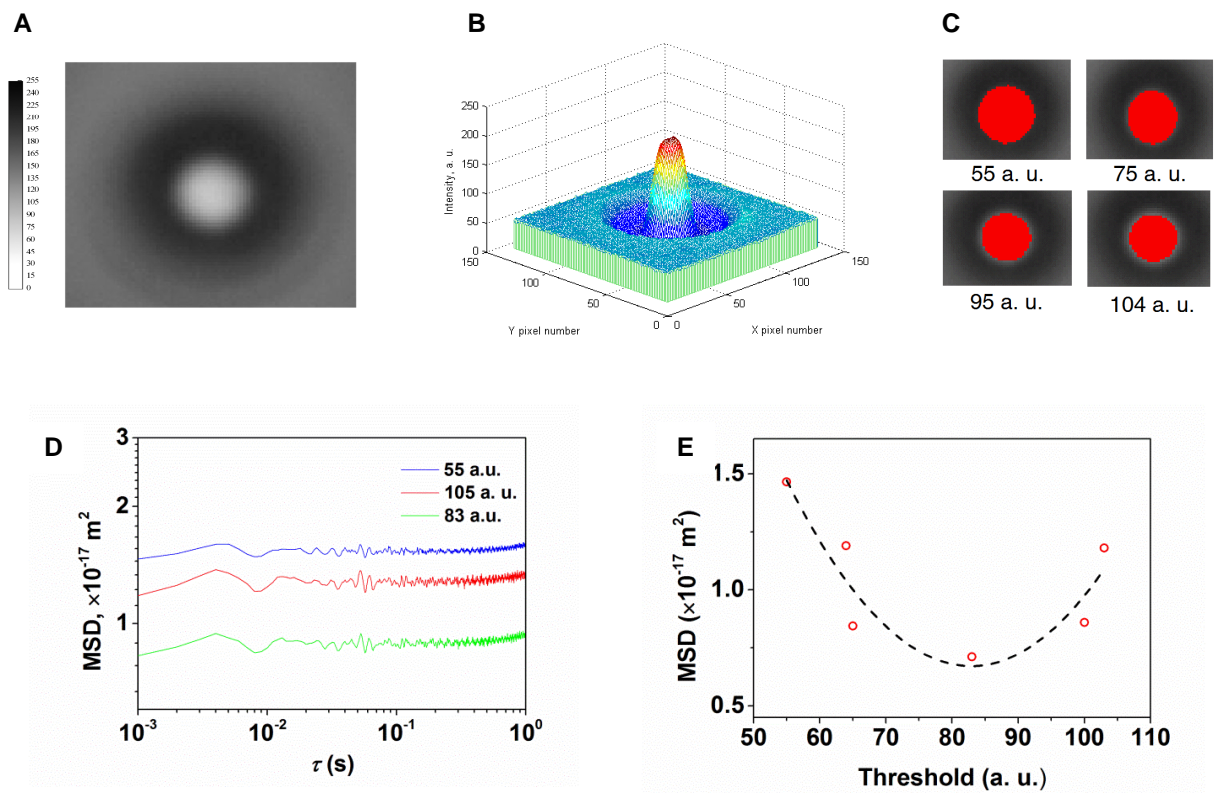

**Fig. S3. Determination of particle's position.** (**A**) Grayscale image of 5 μm particle glued to the bottom substrate of the cell. (**B**) Intensity profile of the particle image. (**C**) Images of the particle recorded at different intensity threshold levels. Red area consists of pixels that are used to calculate the particle's center coordinates. (**D**) MSD of a glued particle at different levels of threshold intensity. (**E**) The optimum value of intensity threshold is determined from the minimum of MSD for the glued particle.

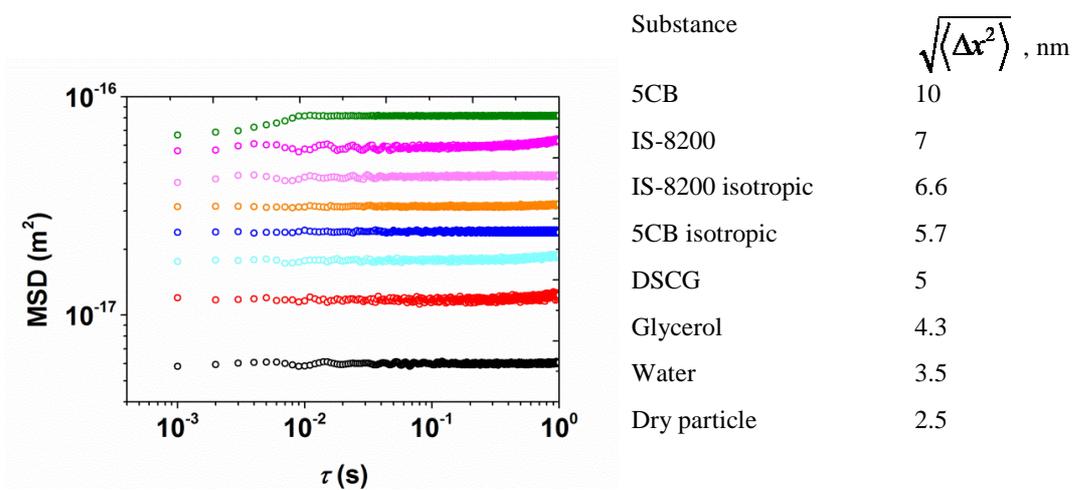

| Substance | $\sqrt{\langle \Delta x^2 \rangle}$, nm |
|---|---|
| 5CB | 10 |
| IS-8200 | 7 |
| IS-8200 isotropic | 6.6 |
| 5CB isotropic | 5.7 |
| DSCG | 5 |
| Glycerol | 4.3 |
| Water | 3.5 |
| Dry particle | 2.5 |

**Fig. S4. Apparent displacement of particle.** Apparent MSD versus time lag for immobilized silica spheres of diameter 5 μm in five different cells (all of thickness 50 µm) filled with three types of a liquid crystals, water, glycerol. Data for an empty cell are labelled as "dry particle". Normal surface anchoring. Higher birefringence results in a higher apparent displacements.

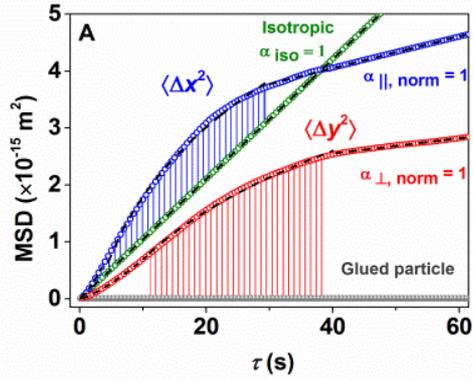
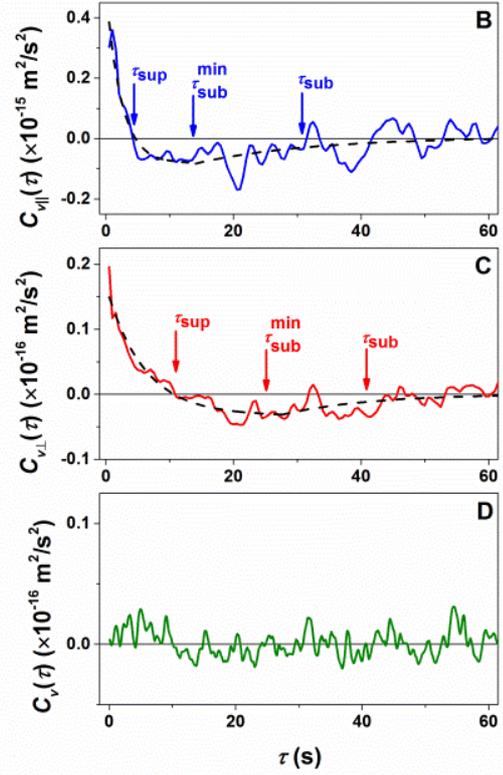

**Fig. S5. MSDs and velocity autocorrelation functions of 5 μm silica spheres with tangential surface anchoring in IS-8200.** (**A**) MSD versus time lag for tangentially anchored sphere in the isotropic ($T = 60°C$) and nematic ($T = 50°C$) phases of IS-8200, in the direction parallel ($x$) and perpendicular ($y$) to the overall director $\hat{\mathbf{n}}_0$. The bottom curve represents apparent MSD of a particle glued to the cell substrate. Cell thickness 50 μm. (**B**) Velocity autocorrelation function for the tangentially anchored sphere moving parallel to $\hat{\mathbf{n}}_0$ and (**C**) perpendicular to $\hat{\mathbf{n}}_0$. (**D**) The same for the isotropic phase. Dashed lines in (**A**) are least-square fits.

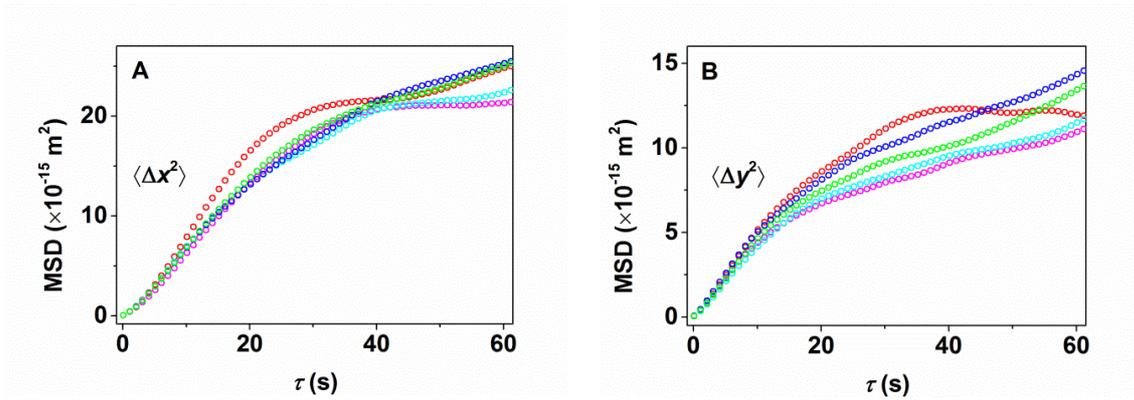

**Fig. S6. MSD vs time lag dependencies for different samples.** MSD versus time lag for five silica spheres of diameter 5 μm in five different cells (all of thickness 50 µm). Normal surface anchoring. The particles are diffusing in the nematic ($T = 50°C$) phase of IS-8200. The MSD is measured in the directions parallel (**A**) and perpendicular (**B**) to the director.

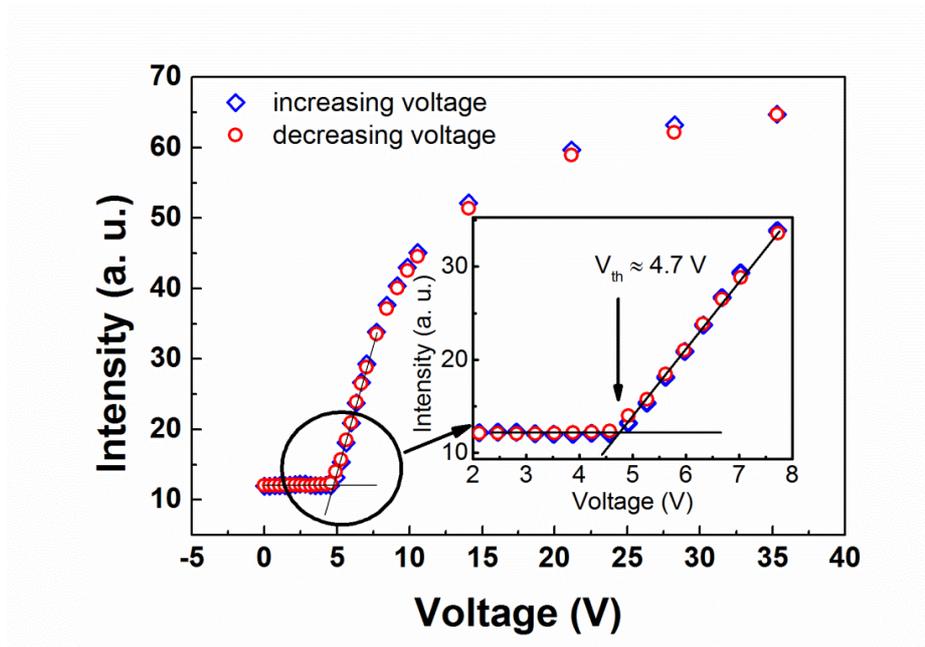

**Fig. S7. Electric field induced Frederiks bend transition in IS-8200.** Intensity of transmitted light as a function of applied voltage. Cell with homeotropic alignment, crossed polarizers. Cell thickness 20 µm.

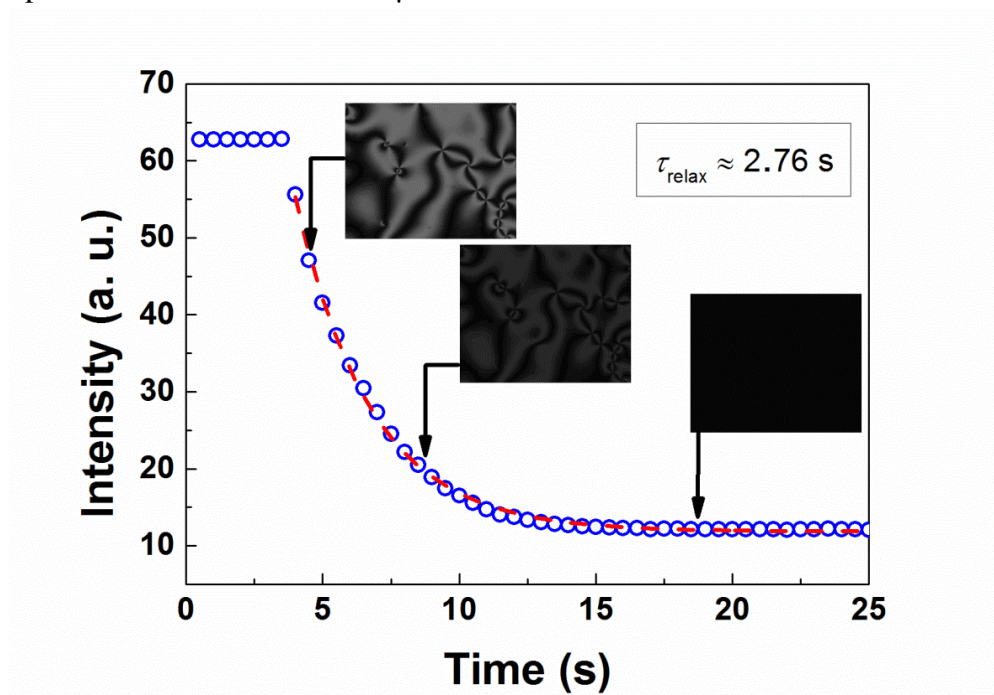

**Fig. S8. Relaxation time of bend deformation in IS-8200.** Intensity of transmitted light as a function of time after voltage is switched off. Homeotropic cell of thickness 20 µm, crossed polarizers.

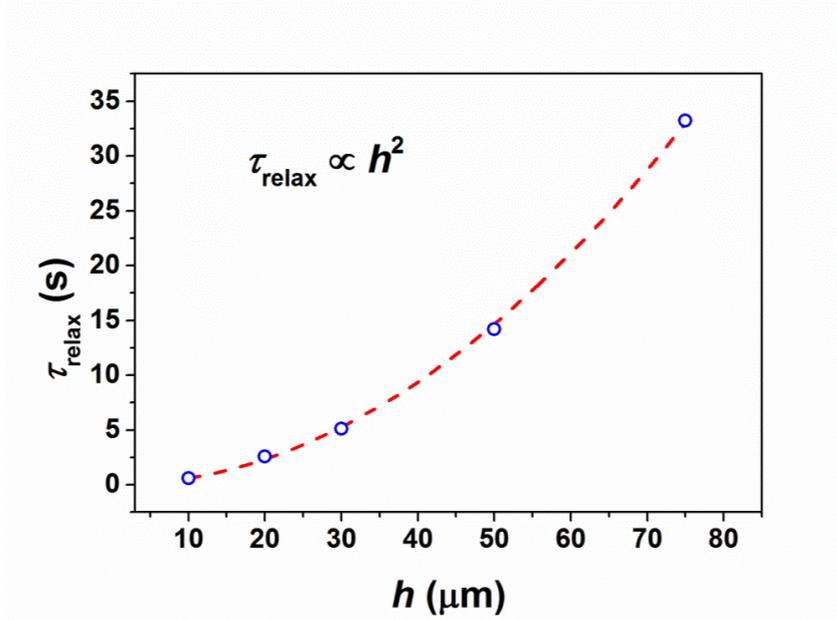

**Fig. S9. Scaling of relaxation time of bend deformation.** Relaxation time versus cell thickness. IS-8200 in a homeotropic cell.

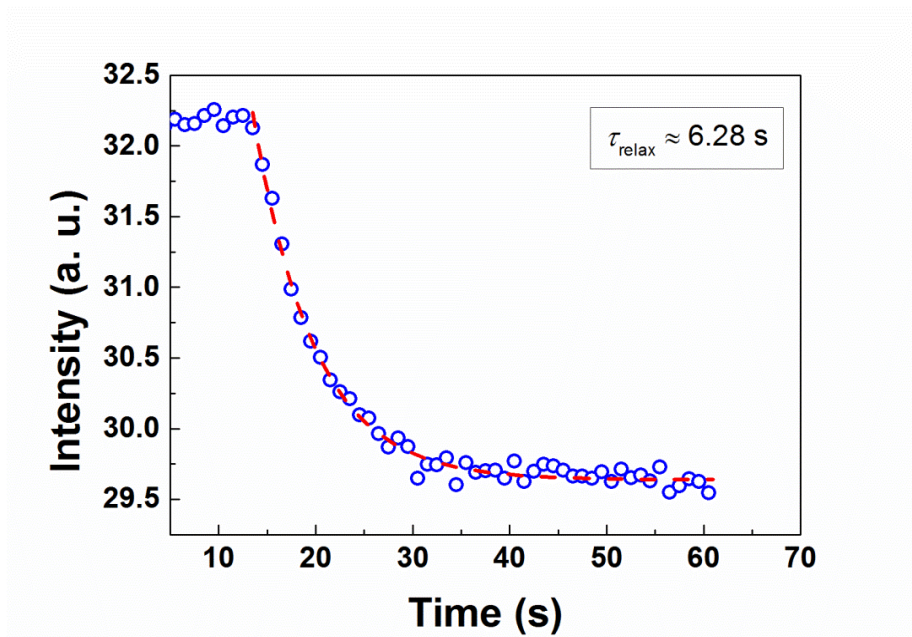

**Fig. S10. Relaxation time of twist deformation in IS-8200.** Dynamics of light transmittance for a relaxation of twist deformation. IS-8200 in a planar cell of thickness 50 µm. Crossed polarizers.

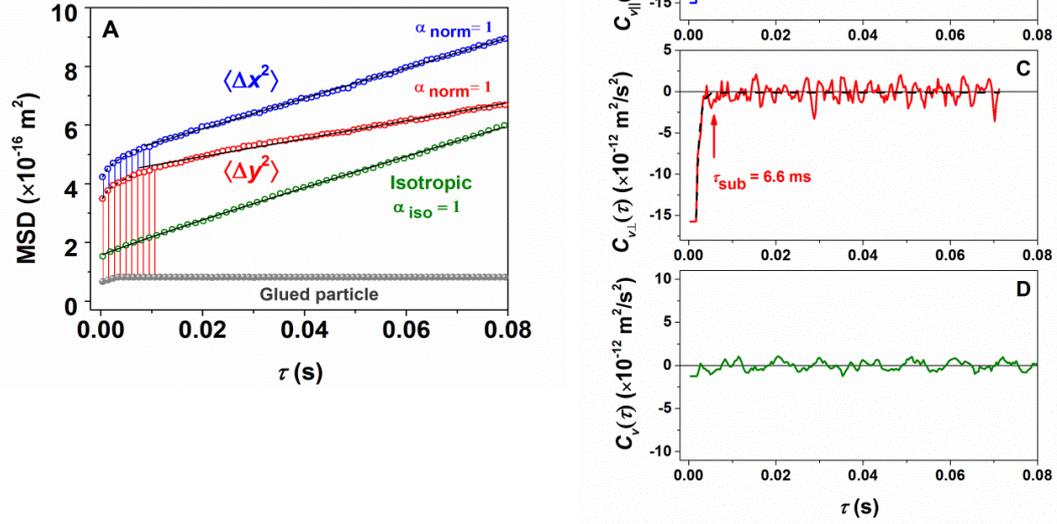

**Fig. S11. MSDs and velocity autocorrelation functions of normally anchored silica sphere of diameter 5 μm diffusing in 5CB.** (**A**) MSD vs time lag for colloidal particles (diameter 5 μm) diffusing in 5CB, in the directions parallel ($x$) and perpendicular ($y$) to the director. The bottom curve represents apparent MSD of immobilized particles in nematic phase. Cell thickness 50 μm. (**B**) Correlation function for the normally anchored sphere moving in the nematic parallel and (**C**) perpendicular to the director. (**D**) Correlation function for the isotropic phase of 5CB.

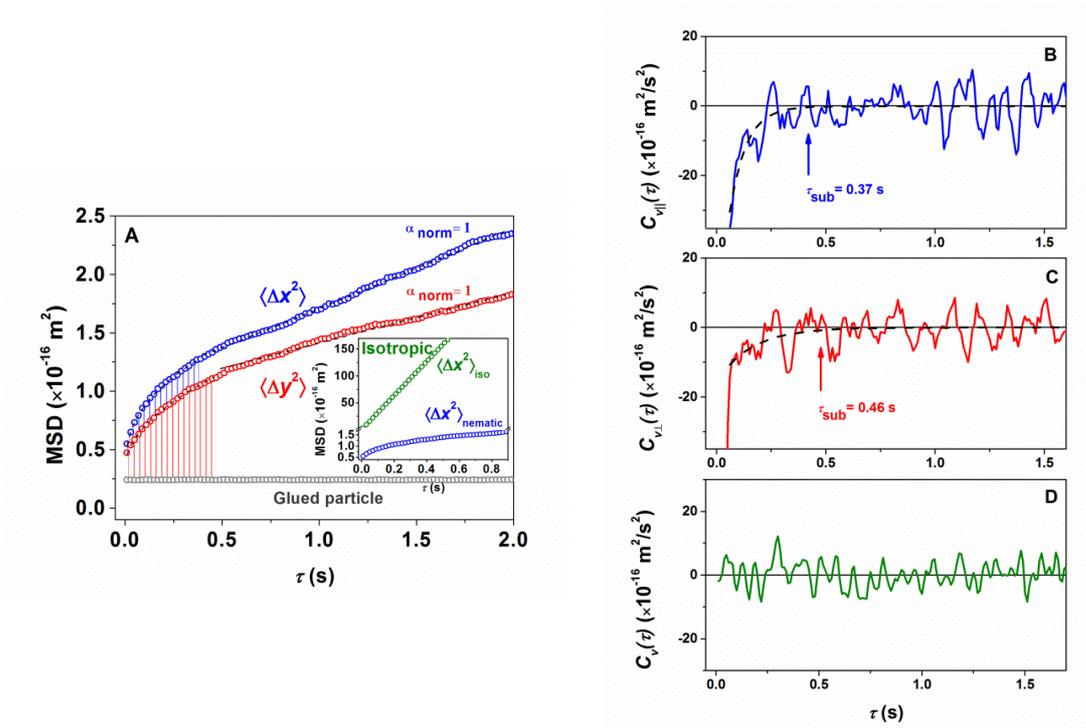

**Fig. S12. MSDs and velocity autocorrelation functions of tangentially anchored silica spheres diffusing in DSCG.** (**A**) MSD vs time lag for colloidal particles (diameter 5 μm) diffusing in DSCG, in the directions parallel ($x$) and perpendicular ($y$) to the overall director. The bottom curve represents apparent MSD of immobilized particles in the nematic phase. Cell thickness 50 µm. Inset: $\langle \Delta x^2(\tau) \rangle$ for nematic ($22°C$) and isotropic ($40°C$) phase. (**B**) Correlation function for the normally anchored sphere moving in the nematic parallel to the overal director and (**C**) perpendicular to it. (**D**) Correlation function for the isotropic phase of DSCG.

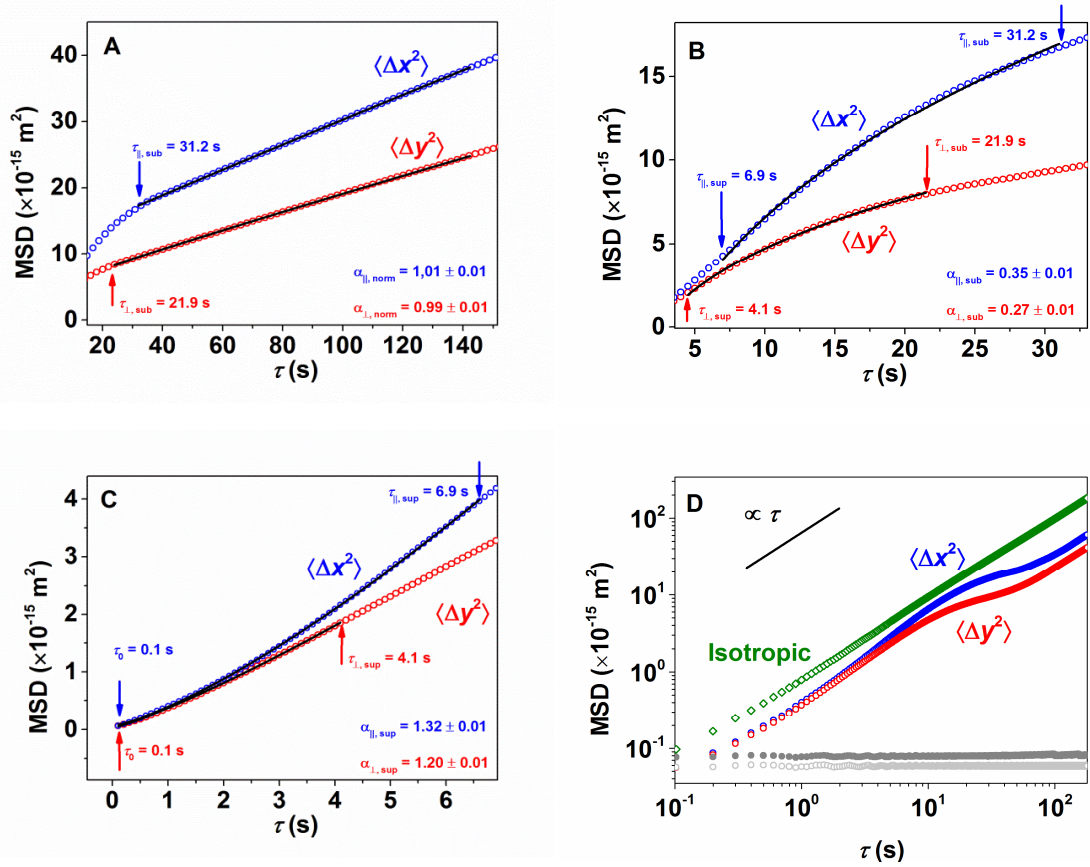

**Fig. S13. MSD vs time lag for 5 µm silica sphere (normal anchoring) diffusing in IS-8200.** (**A**) normal regime, (**B**) subdiffusive regime; (**C**) superdiffusive regime; (**D**) the same data for all three regimes plotted as a single MSD vs. time lag dependence in log-log coordinates. Solid lines in (**A**-**C**) are power law fits. The two bottom curves in (**D**) represent apparent MSDs of an immobilazed particle in the isotropic (open circles) and nematic (filled circles) phases.

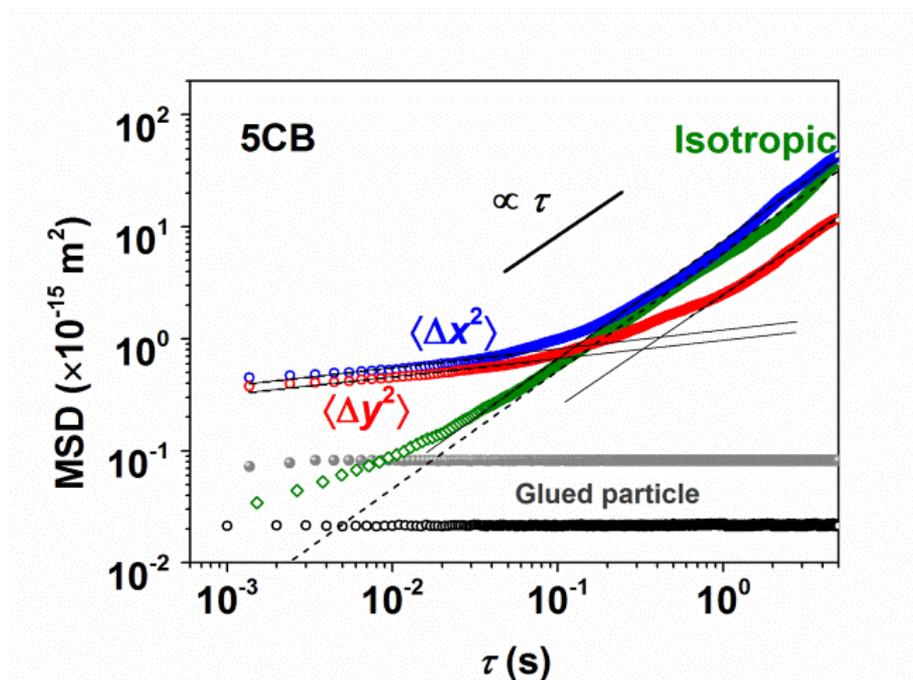

**Fig. S14. MSD vs time lag for 5 μm silica sphere (normal anchoring) diffusing in 5CB plotted in log-log coordinates.** The two bottom curves represent apparent MSDs of an immobilazed particle in the isotropic (open circles) and nematic (grey filled circles) phases.

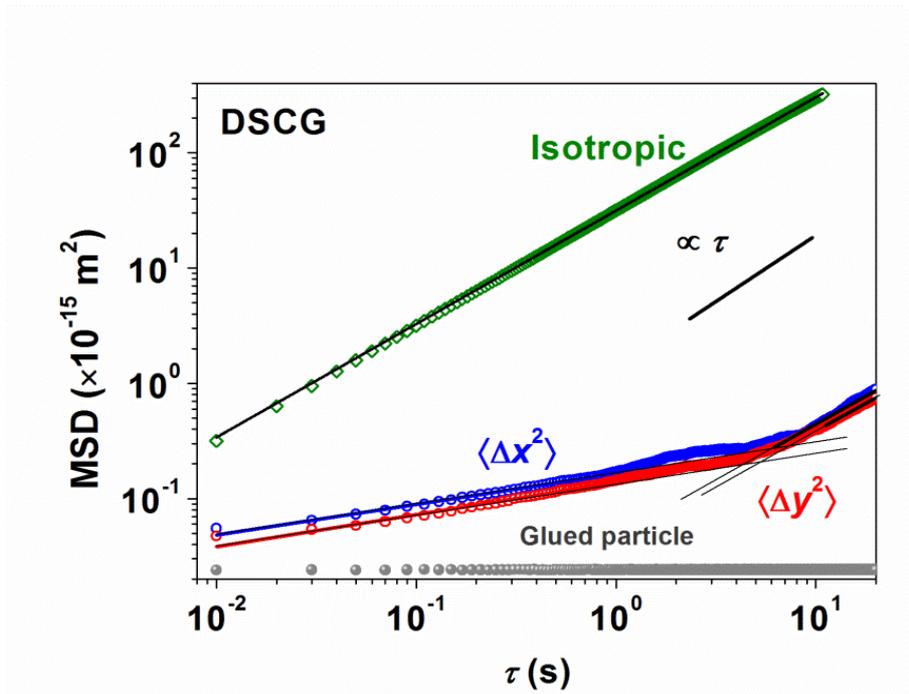

**Fig. S15. MSD vs time lag for 5 μm silica sphere diffusing in lyotropic chromonic liquid crystal DSCG.**

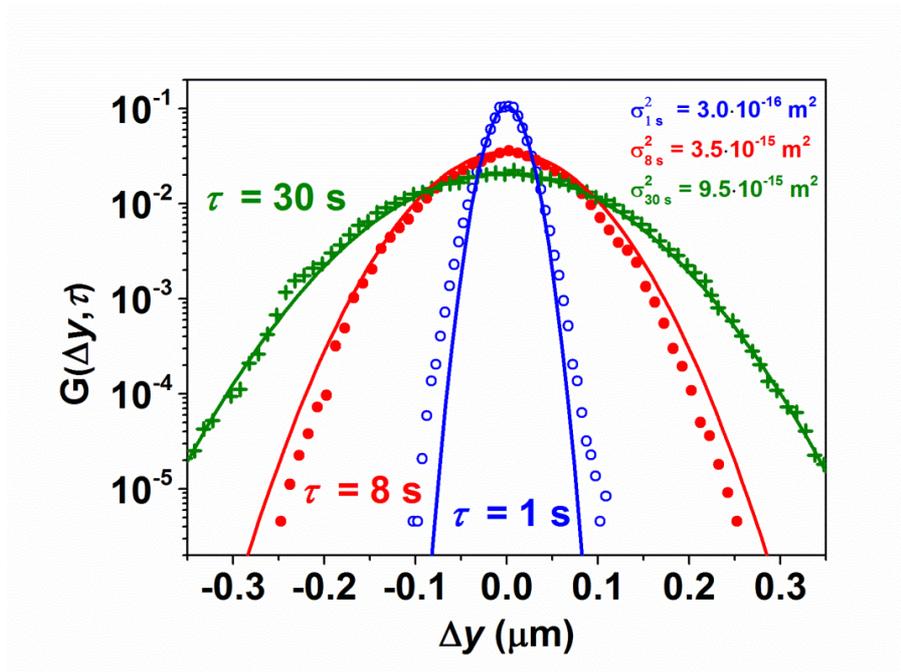

**Fig. S16. Probability distribution of particle displacement perpendicular to the director $\hat{n}_0$, for different time lags, 1 s, 8 s, 30 s.** 5 μm silica sphere in IS-8200, perpendicular anchoring, $T = 50°C$. The solid lines are Gaussian fits. The data obtained from a sample of 200,000 single particle's trajectory steps.

**Supporting table**

**Table S1. Physical properties of nematic liquid crystal IS-8200**

| Birefringence, $\Delta n$ $(\lambda = 510$ nm$)$ | 0.0015 |
|---|---|
| Dielectric anisotropy, $\Delta \varepsilon$ | -1.5 |
| Rotational viscosity, $\gamma_1$ | 2478 mPa |
| Phase diagram | $Cr \xrightarrow{47.0°C} N \xrightarrow{52.3°C} I$ |

**Table S2. Characteristic times and $\alpha_{\parallel,\perp}$ for different diffusion regimes for $d$ = 5 μm silica spheres of the different anchoring conditions dispersed in IS-8200**

| Type of director orientation around the sphere | Superdiffusion | | Subdiffusion | |
|---|---|---|---|---|
| | $\parallel$ | $\perp$ | $\parallel$ | $\perp$ |
| Dipolar structure, normal anchoring | $\alpha_\parallel = 1.2$ $\tau_{sup}^\parallel = 6.9$ s | $\alpha_\perp = 1.3$ $\tau_{sup}^\perp = 4.2$ s | $\alpha_\parallel = 0.4$ $\tau_{sub}^\parallel = 31.2$ s $\tau_{sub}^{min,\parallel} = 14.1$ s | $\alpha_\perp = 0.3$ $\tau_{sub}^\perp = 21.9$ s $\tau_{sub}^{min,\perp} = 9.7$ s |
| Quadrupolar structure, tangential anchoring | $\alpha_\parallel = 1.2$ $\tau_{sup}^\parallel = 4.1$ s | $\alpha_\perp = 1.3$ $\tau_{sup}^\perp = 10.2$ s | $\alpha_\parallel = 0.3$ $\tau_{sub}^\parallel = 29.3$ s $\tau_{sub}^{min,\parallel} = 13.7$ s | $\alpha_\perp = 0.3$ $\tau_{sub}^\perp = 42.7$ s $\tau_{sub}^{min,\perp} = 25.2$ s |

**References and Notes**

1. W. T. Coffey, Y. P. Kalmykov, J. T. Waldron, *The Langevin Equation: With Applications in Physics, Chemistry and Electrical Engineering* Series in Contemporary Chemical Physics (World Scientific, Singapore, 1996), pp. 413.
2. A. Einstein, *Ann. Phys. (Leipzig)* **17**, 549 (1905).
3. J. Sprakel, J. van der Gucht, M. A. C. Stuart, N. A. M. Besseling, *Phys Rev E* **77**, 061502 (2008).
4. I. Y. Wong et al., *Phys Rev Lett* **92**, 178101 (2004).
5. M. M. Alam, R. Mezzenga, *Langmuir* **27**, 6171-6178 (2011).
6. X. L. Wu, A. Libchaber, *Phys Rev Lett* **84**, 3017 (2000) and X. L. Wu, A. Libchaber, *Phys Rev Lett* **86**, 557 (2001).
7. A. Ott, J. P. Bouchaud, D. Langevin, W. Urbach, *Phys Rev Lett* **65**, 2201-2204 (1990).
8. Y. Gambin, G. Massiera, L. Ramos, C. Ligoure, W. Urbach, *Phys Rev Lett* **94**, 110602 (2005).
9. R. Ganapathy, A. K. Sood, S. Ramaswamy, *Europhys Lett* **77**, 18007 (2007).
10. R. Angelico, A. Ceglie, U. Olsson, G. Palazzo, L. Ambrosone, *Phys Rev E* **74**, 031403 (2006).
11. T. G. Mason, D. A. Weitz, *Phys. Rev. Lett.* 7**4**, 1250-1253 (1995).
12. R. W. Ruhwandl, E. M. Terentjev, *Phys Rev E* **54**, 5204-5210 (1996).
13. H. Stark, D. Ventzki, M. Reichert, *J Phys-Condens Mat* **15**, S191-S196 (2003).
14. J. C. Loudet, P. Hanusse, P. Poulin, *Science* **306**, 1525 (2004).
15. G. M. Koenig et al., *Nano Lett* **9**, 2794-2801 (2009).
16. M. Skarabot, I. Musevic, *Soft Matter* **6**, 5476-5481 (2010).
17. J. A. Moreno-Razo et al., *Soft Matter* **7**, 6828-6835 (2011).
18. F. Mondiot et al., *Phys Rev E* **86**, 010401 (2012).
19. D. Abras, G. Pranami, N. L. Abbott, *Soft Matter* **8**, 2026-2035 (2012).
20. M. Pumpa and F. Cichos, *J. Phys. Chem. B* **116**, 14487-14493 (2012).
21. See Supplementary Materials.
22. P. Poulin, H. Stark, T. C. Lubensky, D. A. Weitz, *Science* **275**, 1770-1773 (1997).
23. O. P. Pishnyak, S. Tang, J. R. Kelly, S. V. Shiyanovskii, O. D. Lavrentovich, *Phys Rev Lett* **99**, 127802 (2007).
24. V. Reiffenrath and M. Bremer, Anisotropic Organic Materials, (American Chemical Society, Washington, DC, 2001), chap. 14, pp. 195-205.
25. V. M. Kenkre, R. Kühne, and P. Reineker, *Z Phys B – Condenced Matter* **41**, 177-180 (1981).
26. H. Scher and M. Lax, *Phys Rev B* **7**, 4491-4502 (1973).
27. D. S. Martin, M. B. Forstner, J. A. Kas, *Biophys J* **83**, 2109-2117 (2002).
28. N. Kumar, U. Harbola, K. Lindenberg, *Phys Rev E* **82**, 021101 (2010).
29. P. G. de Gennes, J. Prost, *The Physics of Liquid Crystals*. (Clarendon Press, Oxford, 1993), pp. 598.
30. T. C. Lubensky, D. Pettey, N. Currier, H. Stark, *Phys Rev E* **57**, 610-625 (1998).
31. D. Voloshchenko, O.P. Pishnyak, S.V. Shiyanovskii, O.D. Lavrentovich, *Phys. Rev E* **65,** 060701 (2002).
32. I. Muševič, M. Škarabot, U. Tkalec, M. Ravnik, and S. Žumer, *Science* **313**, 954 (2006).
33. J. C. Crocker, D. G. Grier, *J Colloid Interf Sci* **179**, 298 (1996).
34. A. Kasper, E. Bartsch, H. Sillescu, *Langmuir* **14**, 5004 (1998).
35. M. Cui, J.R. Kelly, *Mol Cryst Liq Cryst* **331**, 49 (1999).

36. H. Stark and D. Ventzki, *Phys Rev E* **64**, 031711 (2001).
37. J. Lydon, *Liq Cryst* **38**, 1663 (2011).
38. Yu. A. Nastishin et al., *Phys Rev E* **72**, 041711 (2005).
39. S. Zhou et al., *Phys Rev Lett* **109**, 037801 (2012).

We thank R. Kamien, E. I. Kats, T. C. Lubensky, E. Weeks for fruitful discussions and O. Boiko, U. Ognysta, A. Nych for help in the experiments. This work was supported by NSF grants DMR 1104850 and 1121288, STCU Project No. 5258, DFFD F35/534-2011, NASU 1.4.1B/10.